\documentclass[rnaas]{aastex62}
\usepackage{amsmath}

\begin{document}

\title{Pre-MAP Search for Transiting Objects Orbiting White Dwarfs}

\correspondingauthor{Brett M. Morris}
\email{bmmorris@uw.edu}

\author{Aislynn Wallach}
\altaffiliation{Pre-Major in Astronomy Program (Pre-MAP) at the University of Washington}

\author[0000-0003-2528-3409]{Brett M. Morris}

\author{Doug Branton}
\altaffiliation{Pre-Major in Astronomy Program (Pre-MAP) at the University of Washington}

\author{Teagan O'Reilly}
\altaffiliation{Pre-Major in Astronomy Program (Pre-MAP) at the University of Washington}

\author{Brittany Platt}
\altaffiliation{Pre-Major in Astronomy Program (Pre-MAP) at the University of Washington}

\author{Ada Beale}
\altaffiliation{Pre-Major in Astronomy Program (Pre-MAP) at the University of Washington}

\author{Andrew Yetter}
\altaffiliation{Pre-Major in Astronomy Program (Pre-MAP) at the University of Washington}

\author{Katie Reil}
\altaffiliation{Pre-Major in Astronomy Program (Pre-MAP) at the University of Washington}

\author{Kristen Garofali}

\author{Eric Agol}
\altaffiliation{Guggenheim Fellow}

\collaboration{(The SPAMS Collaboration)} 
\altaffiliation{Pre-Major in Astronomy Program (Pre-MAP) at the University of Washington}
\affiliation{Astronomy Department, University of Washington, Seattle, WA 98195, USA}

\object{G226-29, Wolf 1516, Wolf 28, SDSS J160401.31+083109.0, SDSS J1152+0248}
\section{Introduction} 

Metal pollution in white dwarf (WD) atmospheres may be the accreted remnants of planetary objects \citep{Jura2009, Gansicke2012, Farihi2013, Veras2015}. After the discovery of disintegrating planetary objects transiting WD 1145+017 \citep{Vanderburg2015}, undergraduates in the University of Washington's Pre-Major in Astronomy Program (Pre-MAP) were inspired to collect photometry of the brightest white dwarfs to hunt for similar transiting objects around other metal-polluted WDs. Prior surveys have yet to make a detection of a transiting planet orbiting a WD \citep{Faedi2011,Fulton2014,Maoz2015,Belardi2016,Sandhaus2016,vanSluijs2018}, yet WDs are still an attractive target for searches of small, rocky planetary material. Since a typical white dwarf is Earth-sized, transits of Earth-sized planets could have depths $>50\%$, so even low S/N photometry has a chance at discovering transiting material.

Rocky planetary material near the Roche limit of a white dwarf host will become tidally disrupted at orbital periods of only 3-10 hours \citep{Veras2014}, enabling the discovery of such transiting objects and their periods in a single night of observations (for a fortuitous observer). The probability of transit for a Moon-sized object at these periods is of order $p_\mathrm{trans} = 0.01$ \citep{Agol2011}.

\section{Observations}
We identified bright, northern, metal-polluted white dwarfs in the SDSS DR10 white dwarf catalog from \citep{Kepler2015} for photometric monitoring. We identified observable WDs from the Astrophysical Research Consortium Small Aperture Telescope (ARCSAT) 0.5-meter telescope at Apache Point Observatory with the observation planning package \texttt{astroplan} \citep{astroplan}. We observed each WD with typical exposure times of 60 seconds in the SDSS g filter with 3x3 binning to maximize observing efficiency. We present here five white dwarf light curves at 60 second cadence: G226-29, Wolf 1516, Wolf 28, SDSS J160401.31+083109.0, and known eclipsing white dwarf binary system SDSS J1152+0248 \citep{Hallakoun2016}.

\section{Results}

\begin{figure*}
\centering
\includegraphics[scale=0.44]{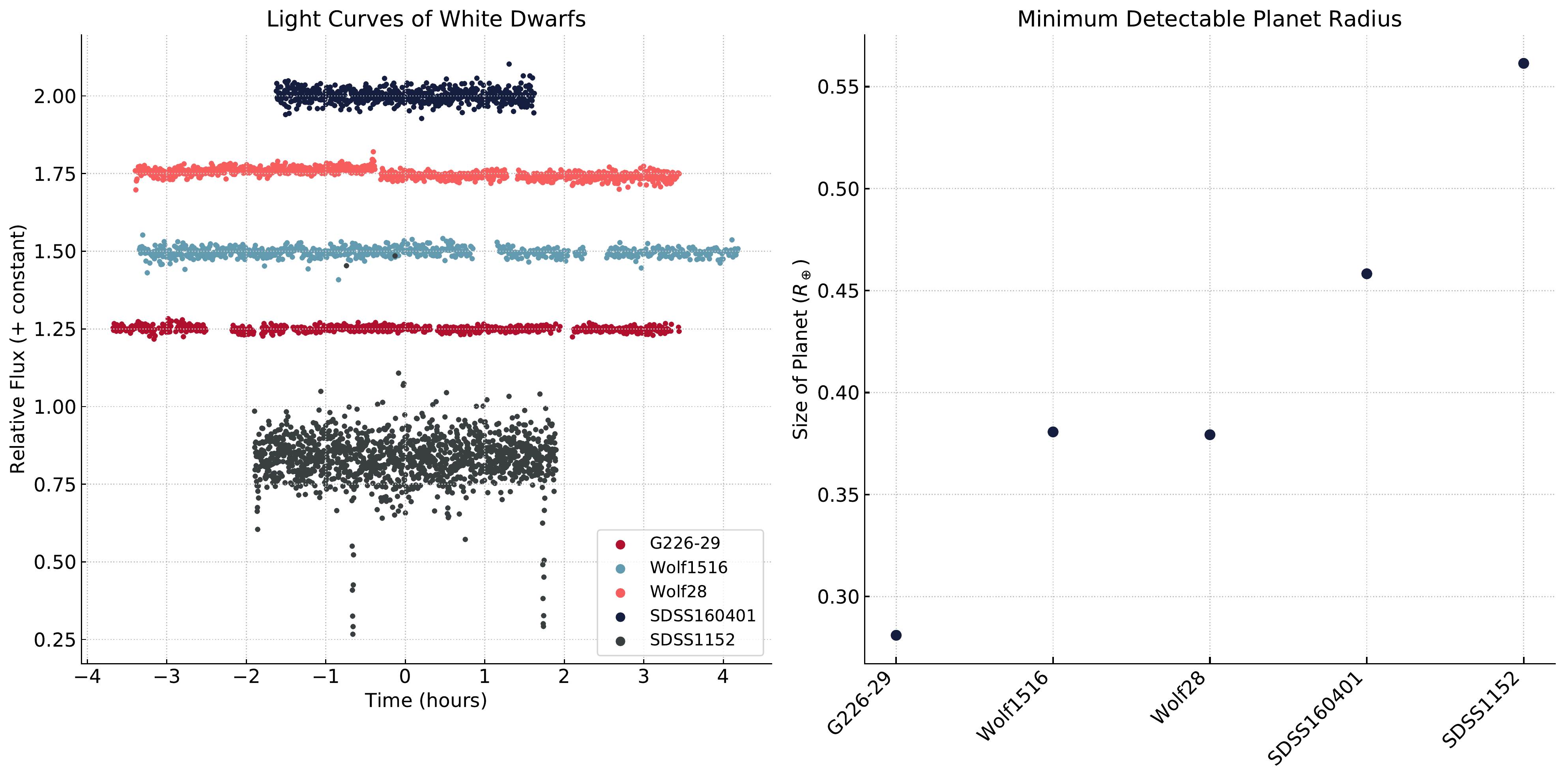}
\caption{ \textsl{Left}: ARCSAT light curves of white dwarfs. \textsl{Right}: Minimum transiting object radius detectable with ARCSAT.}
\label{fig:1}
\end{figure*}

Figure \ref{fig:1} shows photometry of four bright, metal-polluted WDs. Four of these light curves present no obvious periodic dimming events consistent with transiting planets or debris on short orbital periods. The fifth was a control target, SDSS J1152+0248, a known eclipsing white dwarf binary.

After visually inspecting the light curves for transient events, we search for periodic transit events using Lomb-Scargle and box-least squares (BLS) periodograms \citep{gatspy, kovacs2002}. All WDs except the control eclipsing system exhibit negligible flux periodicity, as one might expect from the small transit probabilities predicted in \citet{Agol2011}. For SDSS J1152+0248, BLS detects the correct orbital period of 2.4 h, consistent with other observations \citep{Hallakoun2016}. 

We estimate the minimum potential planet radius observable under our conditions for each star by assuming that we could detect any transit with depth $\delta \approx (R_p/R_\star)^2 >5\times$ the rms photometric scatter (Figure \ref{fig:1}, right panel) with an assumed transit duration of one minute and $R_\star = 1.4R_\oplus$. The ARCSAT photometry has sufficient precision to detect Moon-sized objects or larger at short orbital periods, though no such planets were detected for these targets.

Though we have not detected any transiting debris, we look forward to surveys which may find planets orbiting white dwarfs, such as NASA's TESS, ESA's PLATO, and the Evryscope \citep{Ricker2014, plato, Law2015}. 

\acknowledgments

Based on observations obtained with the Apache Point Observatory 3.5 m and 0.5-meter telescopes, which are owned and operated by the Astrophysical Research Consortium. 


\end{document}